\documentclass[prl,twocolumn,showpacs]{revtex4}

\usepackage{graphicx, epsfig}  
\usepackage{dcolumn} 
\def\be{\begin{equation}}
\def\ee{\end{equation}}
\def\bea{\begin{eqnarray}}
\def\eea{\end{eqnarray}}

\begin{document}

\title{Experimentally Constrained Molecular Relaxation: The Case of Glassy GeSe$_2$}

\author{Parthapratim Biswas}
\email{biswas@phy.ohiou.edu}

\author{De Nyago Tafen}
\email{tafende@helios.phy.ohiou.edu}

\author{D.A. Drabold}
\email{drabold@ohio.edu}

\affiliation{Department of Physics and Astronomy, Ohio University, 
Athens OH 45701, USA}


\date{\today}

\begin{abstract}
An ideal atomistic model of a disordered material should contradict no experiments, 
and should also be consistent with accurate force fields (either {\it ab initio}
or empirical). We make significant progress toward jointly satisfying {\it both} 
of these criteria using a hybrid reverse Monte Carlo approach in conjunction 
with approximate first principles molecular dynamics. We illustrate the method by studying 
the complex binary 
glassy material g-GeSe$_2$. By constraining the model to
agree with partial structure factors and  {\it ab initio} 
simulation, we obtain a 647-atom model in close agreement with experiment, including
the first sharp diffraction peak in the static structure factor. We compute the electronic
state densities and compare to photoelectron spectroscopies. The approach is general and flexible.
\end{abstract}

\maketitle

\section{Introduction}
The modeling of complex materials based upon molecular dynamics simulation has been one 
of the remarkable advances in theoretical condensed matter physics. Whether the 
potentials chosen are empirical or {\it ab initio}, remarkable insights have accrued for 
diverse problems in materials physics and beyond.  
There is, however, an unsatisfying point to the logic of MD simulation: it does not make use of all the information available about a material under study -- notably the information implied by experiments. Simulations often cannot achieve agreement with experiment because of short simulation times, small system sizes or inaccuracies in the interactions. Successful prediction of {\it new} properties is more likely for models in agreement with existing data. Imposition of experimental information may be important in phase-separated or other complex materials for which obtaining a suitable starting structure may be difficult, and for which short MD time scales preclude the emergence of such structures in the model.

A different approach to model construction implemented by McGreevy\cite{gereben,walters,mcgreevy1,mcgreevy2} and 
colleagues is the so-called ``Reverse Monte Carlo" (RMC) method. Here, one explicitly sets 
out to make an atomistic model which agrees with experiments. RMC has been widely used to 
model a variety of complex disordered materials. This is accomplished by making Monte Carlo 
moves which drive a structural model toward exact agreement with one or more experiments.  In practice, RMC is the ideal method to explore the {\it range} of configurations which are consistent with experiment(s). Without adequate limitation to a ``physical" subspace of configuration space, it is unlikely to produce a satisfactory model. That is, 
only a subset of RMC models [which match the experiment(s)] is physically realistic (consistent 
with accurate interatomic interactions).  The imposition of topological/chemical bonding constraints 
in RMC can ameliorate this problem, but not remove it entirely\cite{biswas}.  The
mathematical structure of constrained RMC is a constrained
optimization ``traveling salesman" problem. In our previous implementation of constrained RMC
we formed a positive definite (quadratic) cost or ``penalty" function $\xi$, 
which was then minimized (ideally, but not practically) to zero for a structural 
model which exactly satisfies all constraints imposed:
\be
\label{eq-1} 
\xi=\sum_{j=1}^{K} \sum_{i=1}^{M_j} \eta_{i}^j \{F^j_E(Q_i)- F^j_c(Q_i)\}^2  
   + \sum_{l=1}^L \lambda_l P_l  
\ee
\noindent 

where $\eta_{i}^j$ is related to the uncertainty associated with 
the experimental data points, $K$ is the number of experimental data sets employed,
$M_K$ is the number of data for the $K^{th}$ set and $L$ is the number
of additional (non-experimental) constraints included. The quantity $Q$ is the appropriate generalized variable 
associated with experimental data $F(Q)$ and $P_l>0$ is the 
penalty function associated with each additional constraint and $\lambda_l>0$. 
Such ``additional" constraints can be of many different forms (for example, one may impose chemical or topological ordering, or phase separated units within a continuous random network)
The coordinates of atoms are changed according to Monte Carlo moves, which 
is  akin to a simulated annealing minimization of our cost 
function $\xi$. The method is easy to implement, though care must be taken to 
include the minimum number of independent constraints possible to reduce 
the likelihood of getting ``stuck" in spurious minima. We have shown that 
inclusion of suitable constraints leads to models of a-Si much improved 
compared to RMC models using only the structure factor (first term of 
Eq.~\ref{eq-1}) as constraint\cite{biswas}.

As the creation of models of complex materials is a difficult task, it is of obvious advantage to incorporate {\it all} possible information in fabricating the model. We assert that an ideal model of a complex material should (1) be a minimum (metastable or global) of a suitable energy functional faithfully reproducing the structural energetics, (2) should contradict no experiments. When stated in these terms, our criterion seems quite obvious, yet current simulation schemes do not simultaneously accommodate both criteria, but focus only on one or the other.

In this paper, we merge the power of {\it ab initio} molecular simulation with the {\it a priori} information of experiments to create models consistent with experiments and the chemistry implied by accurate interatomic interactions. To obtain joint agreement, we unite MD with the Reverse Monte Carlo (RMC) method. We name the scheme "Experimentally Constrained Molecular Relaxation" (ECMR). One can understand our scheme as a way to ``tune" 
a structural model using MD within the space of {\it experimentally realistic}  
models as defined by RMC. We choose a troublesome and complex material with a 
long experimental and modeling history: g-GeSe$_2$. 

From an algorithmic perspective, our scheme has important advantages. For example, to model a glass like GeSe$_2$ or SiO$_2$ using first principles methods, the method of choice is to form an equilibrated liquid,  use some dissipative dynamics to simulate an (unphysically) rapid quench of the liquid into an arrested phase and finally to relax this to a local energy minimum, presumably at astronomical fictive temperature (high potential energy). Usually some repeated ``annealing" cycles are also used. If one is interested in a glassy phase all the work of forming and equilibrating the liquid is redundant, and it is a pious hope that the arrested liquid will resemble a real glassy phase. Evidently the likelihood for success is strongly affected by topological and chemical similarity of the melt to the physical amorphous phase. If complex ordering ``self-organization" or phase separation occurs in the physical amorphous phase, the short simulations of conventional {\it ab initio} schemes will surely miss these important structural features. In this vein, we have used ECMR to construct models of a-Si with intermediate range order on a nanometer length scale\cite{thorpe} by inclusion of Fluctuation Electron Microscopy\cite{FEM}.   We note that successful techniques do exist to tackle the time-scale problem\cite{art,nebm}, though these do not enable the inclusion of experimental information. Our method is efficient enough to enable the creation of a 647 
atom model of g-GeSe$2$ using only a workstation. The method is at least a factor of five faster than
a comparable quench from the melt simulation, with its inherent limitations.

\section{Method}

\begin{figure}
\includegraphics[width=3.0 in, height=3.0 in, angle=0]{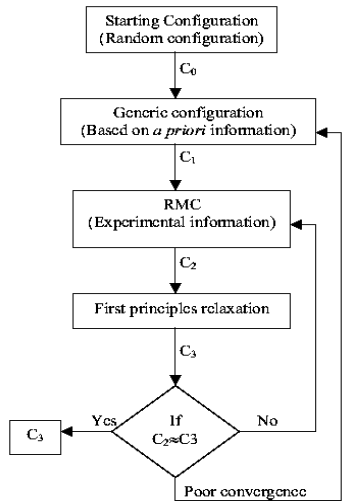}
\caption{ Flow diagram for the ``Experimentally Constrained Molecular Relaxation" method of this
paper.\label{fig1}
}
\end{figure}

The obvious means to incorporate interatomic interactions into an  
RMC simulation is to add a constraint to minimize the magnitude of the force 
on all the atoms according to some energy functional or to minimize the 
total energy. For an 
{\it ab initio} Hamiltonian this is expensive, since Monte Carlo
minimization of Eq.~\ref{eq-1} requires a large number of energy/force calls. 
Thus, we have instead employed a simple ``self consistent" iteration 
scheme (indicated in Fig.~\ref{fig1}): (1) starting with an initial configuration C$_1$, minimize 
$\xi$ to get C$_2$, 2) steepest-descent quench C$_2$ with 
an {\it ab initio} method to get C$_3$, (3) subject the resulting configuration to another 
RMC run (minimize $\xi$ again), repeat steps (2) and (3) until both the MD 
relaxed model and RMC models no longer change with further iteration. 
In this paper, we limit ourselves to the first term in Eq. 1 (the experimental static structure factor), though
additional constraints certainly could be employed. For the
RMC component of the iteration, we make the conventional choice of using Monte Carlo
for the minimization. This is simple and does not require gradients (and thus allows the
use of non-analytic terms in Eq.~\ref{eq-1}~\cite{gradients}, if desired).

We emphasize that our method is {\it flexible}. It's logic suggests that one should include whatever experimental information is available. In this paper we limit ourselves to the pair-correlation functions. In principle, other experiments could be included as well. These might be costly to include (for example to compel agreement with the vibrational density of states, the dynamical matrix would be required at each iteration). The method is equally suited to fast empirical potentials, which would allow studies of very large models. It is also possible to force a close fit to some restricted range of data, and a less precise fit elsewhere if desired. Our scheme also provides insight into the topological signatures of different constraints (experimental or otherwise).  Chemical and or topological constraints could also be maintained as part of the RMC iteration.

Our method can  be understood as a way to minimize an effective potential energy
function $V_{eff}(R) = V(R) + \Lambda \zeta(R)$, where $V(R)$ is the potential energy of the
configuration (denoted by $R$), $\Lambda >0$, and $\zeta$ is a non-negative cost function enforcing experimental (or other)
constraints
as in Eq.~\ref{eq-1}. Empirically, we find that it is possible to find configurations that simultaneously approximately minimize both terms (which implies that the choice of $\Lambda$ is not very important). It is also clear that our method is really {\it statistical}: in general one should generate an ensemble of conformations using ECMR. For adequately large models, self averaging can be expected; in this study of large (647 atom) models of g-GeSe$_2$ we find similar results for two runs; for small systems a proper ensemble average is required. 

The method is new and as such needs to be studied and developed in a number of ways. Nevertheless, we show in this paper that it is relatively easy to model a particularly challenging material with significant advantages in both experimental plausibility of the model and computational efficiency of the algorithm.

\section{Application to glassy GeSe$_2$}

We apply ECMR to glassy GeSe$_2$, a classic glass forming material with challenging physical and technical issues: (1) it displays nanoscale order: a ``first sharp diffraction peak" (FSDP) is observed in neutron diffraction measurement, (2) the packing of GeSe tetrahedra involves both edge- and corner- sharing topologies; (3) the material has interesting photoresponse (understanding of which requires the electronic structure), (4) the material is difficult to simulate with {\it ab initio} techniques\cite{Massobrio 1998,Tafen 2003,Cobb 1996,Zhang 2000}.  The model used in our calculation consists of 647 atoms in a cubic box of size 27.525 {\AA}. 

In the nomenclature of Fig.~\ref{fig1}, C$_1$ is obtained by constraining the coordination 
number (2 for Se, 4 for Ge) and the bond-angle distribution of Se-Ge-Se to an 
approximate Gaussian with an average bond angle 109.5{$^\circ$}. The initial network 
was ``generic" and included none of the detailed local chemistry of Ge and Se aside from 
the coordination and chemical ordering (bond {\it angles} were not constrained in RMC loops). Equal weighting was used for all experimental points in this paper.  Using the method of isotopic substitution, Salmon and Petri~\cite{Petri 2000} were able to separately measure the three (Ge-Se, Ge-Ge and Se-Se) partial structure factors of g-GeSe$_2$. We jointly enforced all three partials (in real space) in the RMC component of the loop in Fig.~\ref{fig1}. The MD relaxation was done with FIREBALL\cite{fireball}. It was found that after the fourth iteration, S(Q) hardly changed.  In Table I, we show the average force per atom at the beginning of each call to MD relaxation; good convergence is observed. Subsequent discussion in this paper is for the last step of the MD, with forces less than $1 \times 10^{-2} eV/\AA$. it was not obvious to us in the beginning that RMC and first principles interatomic interactions could be made ``self-consistent", but for this system at least, reasonable convergence is possible. It is likely that some initial conformations C$_1$ will get ``stuck" and require a new start, but we have not encountered difficulty with this yet.

 \subsection{Structure}
\begin{figure}
\includegraphics[width=4.0 in, height=3.0 in, angle=270]{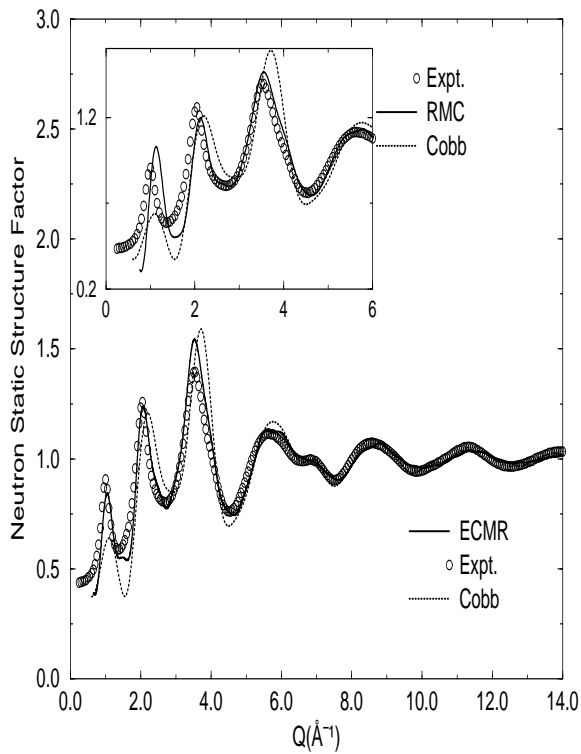}
\caption{
\label{fig2}
Neutron-weighted static structure factor, comparing ECMR model, experiment\cite{Petri 2000} and a quench from the melt made with the same Hamiltonian used with ECMR\cite{Cobb 1996}.  Inset: blowup of small-Q region showing initial RMC model (eg, enforcing experimental structure factor, but without ECMR iterations), experiment\cite{Petri 2000} and quench from the melt model due to Cobb {\it et al.}\cite{Cobb 1996,Zhang 2000}. The first sharp diffraction peak is closely reproduced by ECMR and RMC, and is present but weak in the quenched model.
}
\end{figure}

In Fig.~\ref{fig2}, we compare the RMC, ECMR and experimental structure factors. Here, the RMC model is that obtained by starting with the generic C$_0$ configuration, and forcing agreement on the experimental $S(Q)$ (without any other constraints). While the agreement is very good, it is not perfect. This is to be expected for three reasons: (1) consistency between data and Hamiltonian is never exact; (2) our cell contains 647 atoms, which is compared to the thermodynamic limit and (3) we chose to constrain our model using real space data, which involves Fourier transforms and windowing (this introduces only small errors in this data set).  In Fig. 2 we highlight the differences between experiment\cite{Petri 2000}, a quench from the melt model\cite{Cobb 1996} and the new ECMR model. In the inset of Figure 2, we also illustrate the level of agreement using a pure RMC approach, which is similar to the ECMR result and notably better than quench from the melt. For reference, we have reproduced the full partial structure factors elsewhere\cite{thorpe}.

Note in Fig.~\ref{fig2} that the first sharp diffraction peak (FSDP) is well reproduced, (very close in width and centering,  and much improved from all previous models in height). Moreover, as for our ``Decorate and Relax" (DR) method\cite{tafen}, the large $Q$ structure closely tracks experiment (unlike the experience for quench from the melt models which are too liquid-like and therefore decay too rapidly for large Q). These desirable features are of course ``built in"; we show here that the ECMR method does preserve every important feature of the structure of the glass manifested in S(Q).

An important indicator of network topology and medium range order of GeSe$_2$ glass is the 
presence of edge-sharing and corner-sharing tetrahedra. Raman 
spectroscopy~\cite{Jackson 1999} and neutron diffraction~\cite{Susman 1990} 
studies have indicated that 33\% to 40\% 
of Ge atoms are involved in edge sharing tetrahedra. The 
fraction in our model is found to be 38\%. This was not ``built in" to our modeling,
and is a pleasing prediction arising from the procedure. We also have 
observed that  81\% of Ge atoms in our model are four-fold 
coordinated of which approximately 75\%  form predominant Ge-centered 
structural motifs $Ge(Se_{\frac{1}{2}})_4$ while  6\% are 
ethane-like $Ge_2(Se_{\frac{1}{2}})_6$ units.  The remaining Ge 
atoms are three-fold coordinated and are mostly found to be bonded 
as Ge--Se$_3$ units. On the other hand, the percentage of 
two-, three- and one-fold  coordinated Se atoms are 72\%, 18\% and 10\% respectively. 
M\"ossbauer experiments, where Sn was used as a Ge 
probe~\cite{Boolchand 1982}, estimated that the fraction of Ge 
involved in dimers is 16\% which is again in favor of our model.

\begin{table}
\caption{\label{tab:table1}
The convergence of ECMR described in the text. 
} 
\begin{ruledtabular}
\begin{tabular}{lccccccc}
ECMR iteration & Average force/atom (eV/{\AA}) \\ \hline
1 & $2.242\times 10^{-3}$\\
2 & $7.365\times 10^{-3}$\\
3 & $6.518\times 10^{-4}$\\
4 & $5.019\times 10^{-4}$\\
5 & $4.773\times 10^{-4}$\\
6 & $4.903\times 10^{-4}$\\
7 & $4.686\times 10^{-4}$\\
8 & $4.642\times 10^{-4}$\\
\end{tabular}
\end{ruledtabular}
\end{table} 

\begin{figure}
\includegraphics[width=3.0 in, height=3.0 in, angle=270]{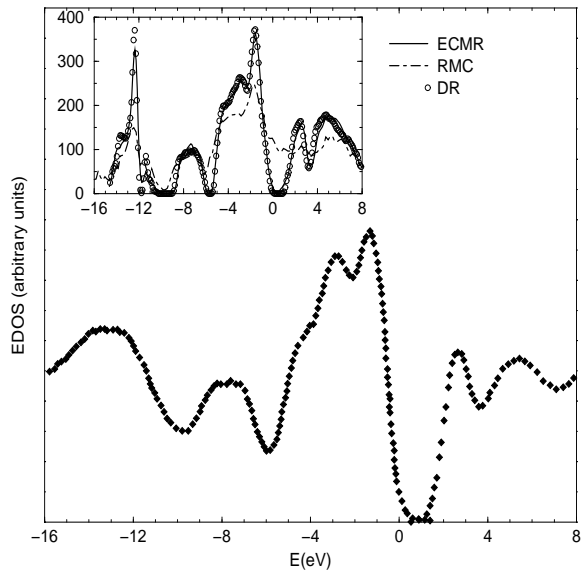}
\caption{
\label{fig3}
The electronic density of states (Gaussian-broadened Kohn-Sham
eigenvalues) for ECMR model of GeSe$_2$, along with the RMC model (not using {\it ab initio} information) and a ``decorate and relax" (DR) model made with the same
Hamiltonian (inset). The XPS\cite{Bergignat 1988} and 
IPES\cite{Hosokawa 1994} data show the occupied (valence 
band) and unoccupied (conduction band) part of the spectrum.  
See Table-II for numerical comparison of the peaks. The Fermi level is at E=0. Both DR and ECMR
reproduce the state density closely, while the RMC model lacks an optical gap.
}
\end{figure}

By integrating partial
radial distribution functions via Fourier transform of structure factors
Petri and Salmon\cite{Petri 2000} obtained
nearest neighbor coordination numbers  $n_{Ge-Ge}$ = 0.25,
$n_{Se-Se}$ = 0.20, and $n_{Ge-Se}$ = 3.7 that corresponds to average
coordination number $\bar n $ = 2.68. The corresponding values from
our model are: $n_{Ge-Ge}$ = 0.17, $n_{Se-Se}$ = 0.30, $n_{Ge-Se}$ = 3.68,
and $\bar n $ = 2.66. The partial and total coordination numbers, therefore,
agree well with experiments (as expected) and are consistent with the 8-N rule which
predicts $\bar n$ = 2.67. The percentage of homopolar bonds present in our
model is found to be about 6.2 \% which is again very close to the value
8 \% noted by Petri and Salmon~\cite{Petri 2000}.

\subsection{Electronic density of states}
Having studied structural properties, we now briefly analyze electronic 
properties of our model. Since structural and electronic properties are 
intimately related, an examination of electronic density 
of states provides an additional test of the validity of the model 
which is derived jointly from structural information and a suitable interatomic interaction. 
The electronic density 
of states (EDOS) is obtained by convolving each energy eigenvalue with 
suitably broadened Gaussian. The ECMR EDOS in the inset of Fig.~\ref{fig3} agrees quite well 
with experimental results obtained from x-ray photo-emission~\cite{Bergignat 1988} 
(XPS), inverse photo-emission spectroscopy~\cite{Hosokawa 1994} (IPES) and 
ultraviolet photo-emission spectroscopy \cite{Hino 1980} (UPS) 
measurements as well as with those obtained in recent theoretical 
studies~\cite{Louie 1982, Pollard 1992, Cobb 1996, Tafen 2003}. 
It is remarkable that the Kohn-Sham eigenvalues (obtained in the Harris approximation)
agree so well with the photoelectron spectroscopy\cite{richard}, particularly as the energy-dependent matrix
element is not included in the calculation.
The substantial splitting 
between the first two peaks of the valence bands named the \textit{$A_1$} 
and \textit{$A_2$} peaks is also well-pronounced. The position of 
the principal peaks obtained from the different models and experiment 
are tabulated in Table II.  The similarity of experiment and theory suggests the utility of
a study of the Kohn-Sham eigenvectors to enable atomistic identification of defects
and bands illustrated in Fig.~\ref{fig3}.

We also compare the EDOS for the RMC model. The RMC model does very poorly, without even showing an optical gap, despite the excellent static structure factor (obtained by construction). By contrast, our DR and the quench from the melt model (not shown) are very close to experiment and ECMR. As the coordination and chemical order is correct in the RMC model, the lack of an optical gap originates in an unrealistic bond angle distribution in the RMC model (something very similar happens in a-Si if only $S(q)$ (and no bond angle constraint)  is used to form the model\cite{biswas}. This result emphasizes the need to compute the density of electron states as an important gauge of the credibility of a model.

\subsection{Vibrations}

It is useful to also examine the vibrational density of states (VDOS) of our 
ECMR model due to the close relationship to its atomic-scale structure and its 
dynamical properties. The VDOS was reported elsewhere 
\cite{Cobb 1996}. Comparing our VDOS with experiment obtained by inelastic 
neutron-scattering \cite{Cappelletti 1995}, the spectrum exhibits the same 
features with somewhat better resolution than
results we reported in Ref. \onlinecite{Cobb 1996}. Three bands can be distinguished: a low 
energy acoustic band involving mainly extended interblock 
vibrations and a high energy optic band consisting of more localized 
intrablock vibrations. The two main bands are clearly separated by the 
tetrahedral breathing (\textit{A$_1$}-\textit{A$_{1c}$}) band. The overall 
agreement is quite reasonable, including a resolved A$_1$ ``companion" mode ``A$_{1c}$".

\begin{figure}
\includegraphics[width=3.0 in, height=3.0 in, angle=270]{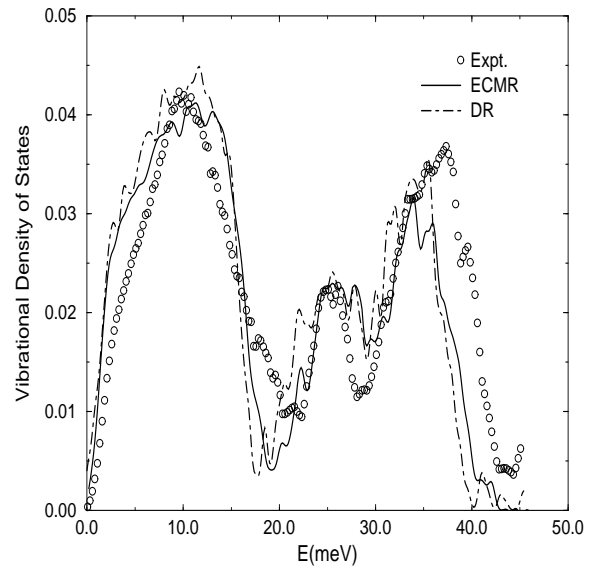}
\caption{
\label{fig4}
Vibrational density of states computed from dynamical matrix for 648 atom models and experiment\cite{Cappelletti 1995}. Nomenclature similar to Fig. 3. }
\end{figure}

In Figure 4, we compare the vibrational density of states of the ECMR model with experiment\cite{Cappelletti 1995} and for completeness our decorate and relax model
including 648 atoms along with the ECMR model. We do not present the RMC result, as the system is not at equilibrium according to FIREBALL, which would therefore lead to many imaginary frequencies in the density of states. 
While generally DR and ECMR are quite similar, we note some difference in  the tetrahedral breathing A$_1$ band (near 25 meV), including a slightly different A$_1$-A$_{1c}$ splitting. This is probably
because the ratio of edge to corner sharing tetrahedra is different ($\approx$ 29\% of Ge atoms are involved in edge sharing tetrahedra in the DR model). This suggests that the VDOS has some sensitivity to medium range order (tetrahedral packing) in this glass.

\begin{table}
\caption{\label{tab:table4}The positions of the \textit{A$_1$}, \textit{A$_2$}, 
\textit{A$_3$} and \textit{B} peaks in the EDOS of GeSe$_2$ glass compared to 
experimental results \cite{Hino 1980}.} \begin{ruledtabular}
\begin{tabular}{lccccccc}
(eV)& \textit{A$_1$} & \textit{A$_2$} & \textit{A$_3$} & \textit{B} \\
\hline
Present work & -1.55 & -3.0 & -4.6 & -7.4 \\
Experiment \cite{Hino 1980} & -1.38 & -3.0 & -4.6 & -7.8 \\
Melt and quench\cite{Cobb 1996} & -1.4 & -2.7 & -4.6 & -7.0 \\
Decorate and relax\cite{Tafen 2003} & -1.36 & -2.8 & -4.5 & -7.2 \\
\end{tabular}
\end{ruledtabular}
\end{table} 

\section{Conclusion} 
In summary, we have proposed a new method which enables the inclusion of {\it a priori} information (experimental or otherwise) into molecular simulation. We have shown that the method is effective for a  challenging material, g-GeSe$_2$.

\section{Acknowledgments}
We thank Dr.\,P.S. Salmon for providing us with experimental data, and 
Professor Himanshu Jain for helpful discussions. We 
acknowledge the support of National Science Foundation for support 
under Grants Nos.  DMR-0074624, DMR-0205858 and DMR-0310933.


\begin{thebibliography}{99}

\bibitem{gereben}
O.~Gereben and L.~Pusztai, Phys.~Rev.~B {\bf 50}, 14 136 (1994).

\bibitem{walters}
J.~K.~Walters and R.~J.~Newport, Phys.~Rev.~B {\bf 53}, 2405 (1996).

\bibitem{mcgreevy1}
R.~L.~McGreevy, J.~Phys.: Condens.~Matter {\bf 13}, R877 (2001).

\bibitem{mcgreevy2}
R.~L.~McGreevy and L.~Pusztai, Mol.~Simul. {\bf 1}, 359 (1988).

\bibitem{biswas}
P.~Biswas, R.~Atta-Fynn, and D.~A.~Drabold, Phys.~Rev.~B {\bf 69}, 
195207 (2004). 

\bibitem{thorpe}
P.~Biswas, D.N. Tafen, R.~Atta-Fynn, and D.~A.~Drabold, J.~Phys.: Condens.~Matter {\bf 16}, XXXX (2004).

\bibitem{FEM}
M. M. J. Treacy, J. M. Gibson, Acta Cryst. A {\bf 52}, 212 (1996).

\bibitem{art}
See for example, G. T. Barkema and N. Mousseau, Phys.~Rev.~B {\bf 62}, 4985 (2000) and references therein.

\bibitem{nebm}
G. Henkelman and H. Jonsson, J. Chem. Phys. {\bf 113}, 9978 (2000);
G. Henkelman, B. P. Uberuaga, and H. Jonsson, J. Chem. Phys. {\bf 113}, 9901 (2000).

\bibitem{gradients} It is possible that suitable gradient-based methods could provide
more rapid convergence, a point we do not investigate here.

\bibitem{Massobrio 1998}C. Massobrio, A. Pasquarello, and R. Car, Phys. Rev. Lett. {\bf 80},
2342 (1998).

\bibitem{Tafen 2003}D.N. Tafen and D.A. Drabold, Phys. Rev. B {\bf 68}, 165208 (2003).

\bibitem{Cobb 1996}M. Cobb, D.A. Drabold, and R.L. Cappelletti, Phys. Rev. B {\bf 54}, 
12162 (1996).

\bibitem{Zhang 2000}X. Zhang and D.A. Drabold, Phys. Rev. B {\bf 62}, 15695 (2000).

\bibitem{Petri 2000}I. Petri and P.S. Salmon, Phys. Rev. Lett. {\bf 84}, 2413 (2000).

\bibitem{fireball}O.F. Sankey and D.J. Niklewski, Phys.~Rev.~B {\bf 40},
3979 (1989); O.F. Sankey, D.A. Drabold, and G.B. Adams, Bull. Am. Phys. Soc. {\bf 36},
924 (1991). 

\bibitem{tafen}D.A. Drabold, Jun Li, and D.N. Tafen, J. Phys. Cond. Matt. {\bf 15},
S1529 (2003).

\bibitem{Jackson 1999}
K. Jackson, A. Briley, S. Grossman D.V. Porezag, and M.R. Pederson, 
Phys. Rev. B {\bf 60}, R14985 (1999). 

\bibitem{Susman 1990}
S. Susman, K.J. Volin, D.G. Montague, and D.L. Price, 
J. Non-Cryst. Solids {\bf 125}, 168 (1990).


\bibitem{Boolchand 1982}P. Boolchand, J. Grothaus, W.J. Bresser, 
and P. Suranyi, Phys. Rev. B {\bf 25}, 2975 (1982).

\bibitem{Bergignat 1988}E. Bergignat, G. Hollinger, H. Chermette, 
P. Pertosa, D. Lohez, M. Lannoo, and M. Bensoussan Phys. Rev. B {\bf 37}, 4506 (1988).

\bibitem{Hosokawa 1994}S. Hosokawa, Y. Hari, I. Ono, K. Nishihara, M. Taniguchi,
O. Matsuda, and K. Murase, J.~Phys.: Condens.~Matter {\bf 6}, L207 (1994).


\bibitem{Hino 1980}S. Hino, T. Takaharshi, and Y. Harada, Solid State 
Communi. {\bf 35}, 379 (1980).

\bibitem{Louie 1982}S.G. Louie, Phys. Rev. B {\bf 26}, 5993 (1982).

\bibitem{Pollard 1992}W. Pollard, J. Non-Cryst. Solids {\bf 144}, 70 (1992).

\bibitem{richard} R. M. Martin, {\it Electronic Structure, Basic Theory and Practical Methods}, Cambridge University Press, Cambridge (2004). page 144.

\bibitem{Cappelletti 1995}R.L. Cappelletti, M. Cobb, D.A. Drabold, and 
W.A. Kamitakahara, Phys. Rev. B {\bf 52}, 9133 (1995).

\end{thebibliography}
\end{document}